\documentclass[useAMS,usenatbib,onecolumn]{mn2e}

\usepackage{dcolumn}
\usepackage{graphicx}
\usepackage{url}
\usepackage{color}

\newcommand{\cm}{cm$^{-1}$}

\newcommand{\ai}{\textit{ab initio}}

\newcommand{\eqref}[1]{(\ref{#1})}

\newcommand{\duo}{{\sc Duo}}

\newcommand{\X}{$X\,{}^1\Sigma^{+}$}
\newcommand{\A}{$A\,{}^1\Sigma^{+}$}
\newcommand{\Ap}{$A^{\prime}\,{}{}^1\Pi$}
\newcommand{\astate}{$a\,{}^3\Pi$}
\newcommand{\bstate}{$b\,{}^3\Sigma^{+}$}
\newcommand{\B}{$B\,{}^1\Pi$}
\newcommand{\Cp}{$C^\prime\,{}^1 \Sigma^+$}
\newcommand{\cstate}{$c\,{}^3\Sigma^{+}$}

\newcommand{\p}{^\prime}

\title[ExoMol XIII: Line list for CaO]{ExoMol molecular line lists  - XIII: The spectrum of CaO}

\date{\today}
\author[Sergei N. Yurchenko, Audra Blissett, Usama Asari,  Marcus Vasilios, Christian Hill and Jonathan Tennyson ]{\large
Sergei N. Yurchenko, Audra Blissett, Usama Asari,  Marcus Vasilios, Christian Hill and Jonathan Tennyson  \\
Department of Physics and Astronomy, University College London, London WC1E 6BT,
UK}
\date{Accepted XXXX. Received XXXX; in original form XXXX}

\pagerange{\pageref{firstpage}--\pageref{lastpage}} \pubyear{2012}
\begin{document}
\maketitle
\begin{abstract}

An accurate line list for calcium oxide is presented covering transitions between all
bound ro-vibronic levels from the five lowest electronic states \X, \Ap,
\A, \astate, and \bstate. The ro-vibronic energies and corresponding
wavefunctionts were obtained by solving the fully coupled Schr\"{o}dinger
equation. \textit{Ab initio} potential energy, spin-orbit, and electronic
angular momentum curves were refined by fitting to the experimental
frequencies and experimentally derived energies available in the
literature. Using our refined model we could (i) reassign the vibronic states for a large portion of the experimentally derived energies
[van Groenendael A., Tudorie M., Focsa C., Pinchemel B., Bernath P. F., 2005, J. Mol. Spectrosc., 234, 255],
(ii) extended this list of energies to $J=79-118$ and (iii) suggest a new description of the resonances from the \A--\X\ system.
We used high level \textit{ab initio} electric dipole moments
reported previously [Khalil H., Brites V., Le Quere F., Leonard C., 2011,
Chem. Phys., 386, 50]  to compute the Einstein A coefficients.  Our work is the first fully coupled description of this system. Our line list is the most complete catalogue of spectroscopic transitions available for $^{40}$Ca$^{16}$O and is applicable for temperatures up to at least 5000~K. CaO has yet to be observed astronomically
but its transitions are characterised by being particularly strong which should
facilitate its detection. The CaO line list is made available in an electronic form as supplementary data to this article and at \url{www.exomol.com}.

\end{abstract}

\label{firstpage}

\section{Introduction}

 The discovery of candidate ``Lava planets'', such as Corot-7b \citep{09LeRoSc.exo}, Kepler-10b \citep{11RoDeDe.exo}
and 55 CnC e \citep{11WiMaDa.exo}, which are thought to
support day-side temperatures around 3000~K, has opened a completely
new field of (exo-)planet spectroscopy. Theoretical studies have proposed possible formation mechanisms
for these planets \citep{11LeOdKu.exo},
compositions for their hot atmospheres \citep{11CaMexx.exo,12ScLoFe.exo}
and their possible spectroscopic signatures \citep{15ItIkKa.exo}.
They are considered important targets for space telescopes wishing to perform spectroscopic
characterisation of their atmospheres \citep{jt523,14SaLeRo.exo,jt606}.

Calcium oxide, which is quicklime in its solid form, is  a possible
constituent of rocky type exoplanets where it should exist in a gaseous form at
higher temperatures.
However,  to the best of our  knowledge the CaO molecule has not yet been detected astronomically
although it has been searched for in molecular clouds by \citet{98SaWhKa.CaO}.
\citet{79HoWiCh.CaO} estimated the flux-range for CaO absorption in
stellar atmospheres and molecular clouds; this analysis has been
extended to the atmosphere of exoplanet CoRoT-7b by \citet{11GuCaEr.CaO} who suggest that the
signature for CaO should be particularly strong.

The first experimental study of the gaseous hot CaO using discharge was
performed by \citet{11Chxxxx.CaO}. \citet{32Brxxxx.CaO} performed an experimental study of the \A--\X\ electronic system of CaO, which was later complemented and reanalyzed in a series of experimental works by Lagerqvist and co-workers
\citep{50HuLaxx.CaO,51HuLaxx.CaO,54Laxxxx.CaO,54LaHux1.CaO,54LaHuxx.CaO,57LaNiBa.AlO} and also in a theoretical work by \citet{74Fixxxx.CaO}. A number of band heads from the \A--\X\ system  were reported experimentally by \citet{68BrHaxx.CaO}.
Recently  the \A--\X\ electronic system was addressed by \citet{05VaTuFo.CaO} in their comprehensive study with the high-resolution
Fourier transform spectroscopy.
The strong \Ap--\X\ band was studied experimentally by
\citet{75FiCaJo.CaO}  with a number of band heads corresponding to vibrationally
excited states reported  and by \citet{00FoPoPi.CaO} where a large number of ro-vibronic transitions were assigned.
\citet{78HoPeCr.CaO} reported millimeter wave spectra of CaO and also
attempted a search of CaO in stars and molecular clouds though without success.
The hot infrared spectrum of CaO (\X\ -- \X) was observed by \citet{88BlHexx.CaO} and \citet{89HeBlxx.CaO}.

The orange arc bands of CaO, which arise from transitions between excited
triplet states were studied by \citet{82MaScGo.CaO} and
\citet{91PlNixx.CaO}. Other experimental works on the electronic spectra of CaO
include studies of the \cstate\ -- \astate\ band using the sub-Doppler intermodulation spectroscopy by \citet{89NoCrSc.CaO},
\B\ -- \bstate\ bands by \citet{90BaFix1.CaO} and the green-band transitions of
the $F\,{}^1\Pi$  -- \Ap\ and \B\ -- \Ap\ systems by \citet{90BaFixx.CaO} using the dispersed laser
fluorescence spectroscopy; the \Cp\ -- \Ap, $e\,{}^3\Sigma^-$ -- \astate\ and
$E\,{}^1\Sigma^-$ -- \Ap\ bands using a combination of laser-induced fluorescence, resolved fluorescence, and
optical double resonance techniques  by \citet{89BaFixx.CaO,90BaNoSo.CaO}. The electronic transition
strengths of the \B\ -- \X\ and \Cp\ -- \X\ bands of  CaO  were determined by absorption measurements in a shock tube by \citet{80SvKuKu.CaO,80SvKuK1.CaO}.

Lifetimes of the \A\ -- \X\ and orange arc bands were measured by
\citet{91PlNixx.CaO}, which provide a useful check for our \ai\ transition dipoles
and hence predicted intensities.
%\red{CAN BE COMPARED TO OUR VALUES, 149+/-11 ns, delta v$<$3}.
Dissociation energies were obtained experimentally by
\cite{64DrVeEx.CaO} using mass spectroscopy and by \cite{80IrDaxx.CaO} from
chemiluminescence. On the \ai\ side,
\citet{11KhBrLe.CaO,12KhLeBr.CaO} reported high level \ai\ studies of CaO including accurate potential energy curves
(PECs), spin-orbit curves (SOCs), electronic angular momentum curves (EAMCs),
dipole moment curves (DMCs), transition dipole moment curves (TDMCs)  as well different structural properties including the dissociation energies. These results provide the input to our nuclear motion calculations and are considered in detail below.

%\red{THIS PARAGRAPH IS ABOUT VERY OLD PAPERS AND CAN BE OMITTED, I THINK}
%\citet{68Yoxxxx.CaO} presented the \ai\ potential energy and dipole moment
%curves for the ground electronic state using MO-approximation.
%\citet{78BaYaxx.CaO} reported SCF energies for a number of electronic states of CaO.
%MCSCF/CI method was used for the electronic structure description of the
%low lying states of CaO by \citet{xxxx} \red{????}.
%\citet{70CaKaMo.CaO} used HF for the low lying triplet states.
%\textit{Ab initio} values of dissociation energies of CaO were also reported by \citet{11KhBrLe.CaO} in their
% CCSD(T) (coupled clusters singles and doubles plus perturbative triples) in
%conjunction with core-valence correlation.

The aim of this work was to produce a molecular line list for calcium oxide,
$^{40}$Ca$^{16}$O (hereafter CaO),  as part of the ExoMol project \cite{jt528}. This line list is a catalogue of transitions required for modelling
absorption/emission of the molecule in question. We use the \ai\ curves by
\citet{11KhBrLe.CaO,12KhLeBr.CaO,12Lexxxx.CaO} covering the lowest five electronic states,
\X, \Ap, \A, \astate, and \bstate. The \ai\ PECs, SOCs, and EAMC were refined
by fitting to the experimental energies and transition frequencies from the
literature. The line list is used to simulate different spectra for a range of
temperatures of CaO which we validate against experiment.

\section{Method}

\begin{figure}
\begin{center}
\includegraphics[width=220pt]{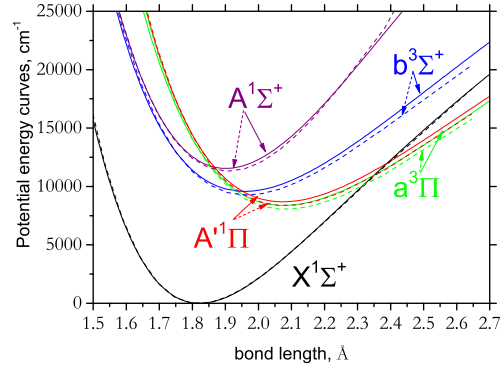}
\caption{ \label{f:PECs}
{\it Ab initio} (dotted curves) and fitted (full curves) potential energy curves for the lowest five electronic
states of CaO. }
\end{center}
\end{figure}

\begin{figure}
\begin{center}
\includegraphics[width=220pt]{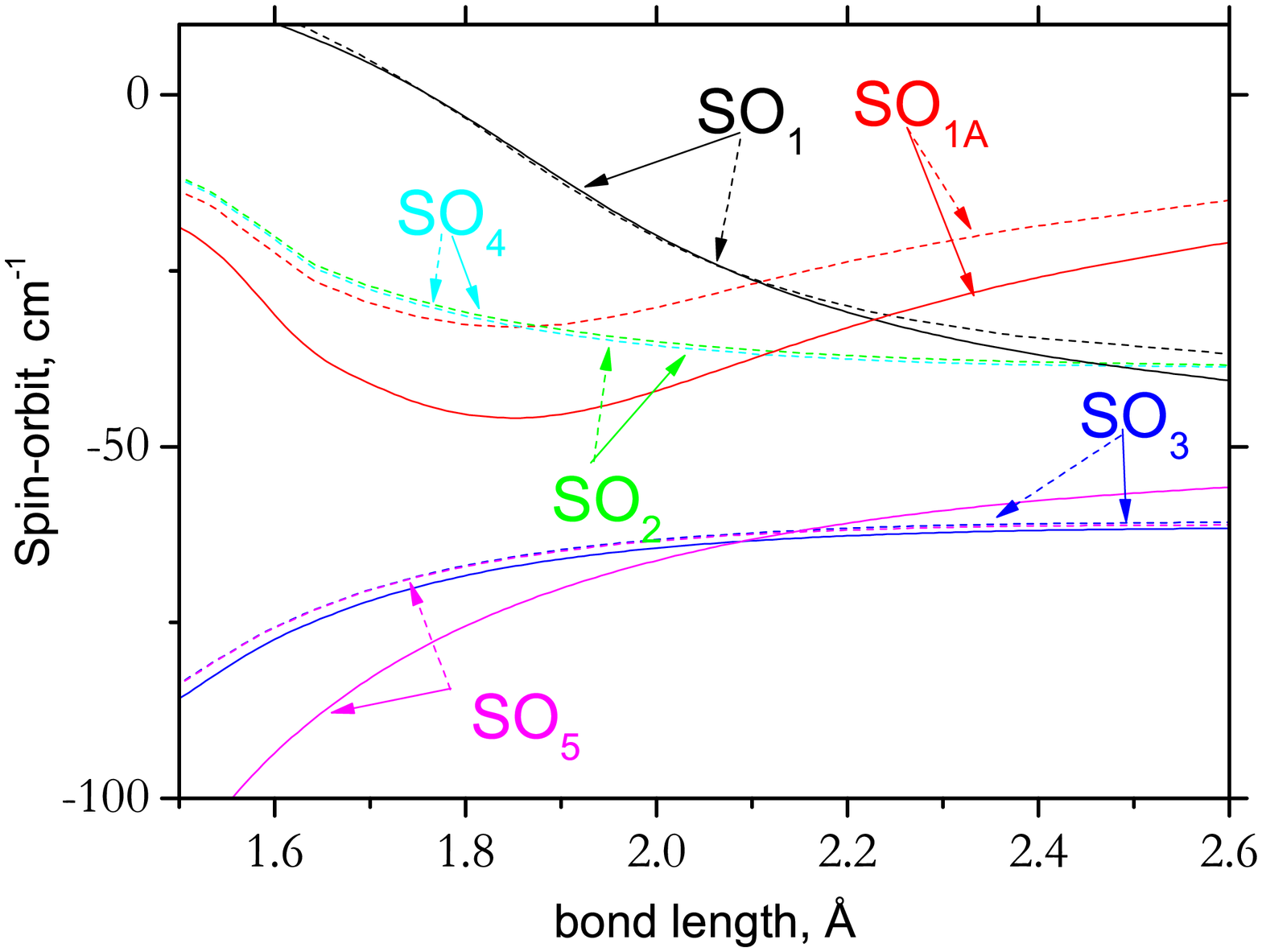}
\includegraphics[width=220pt]{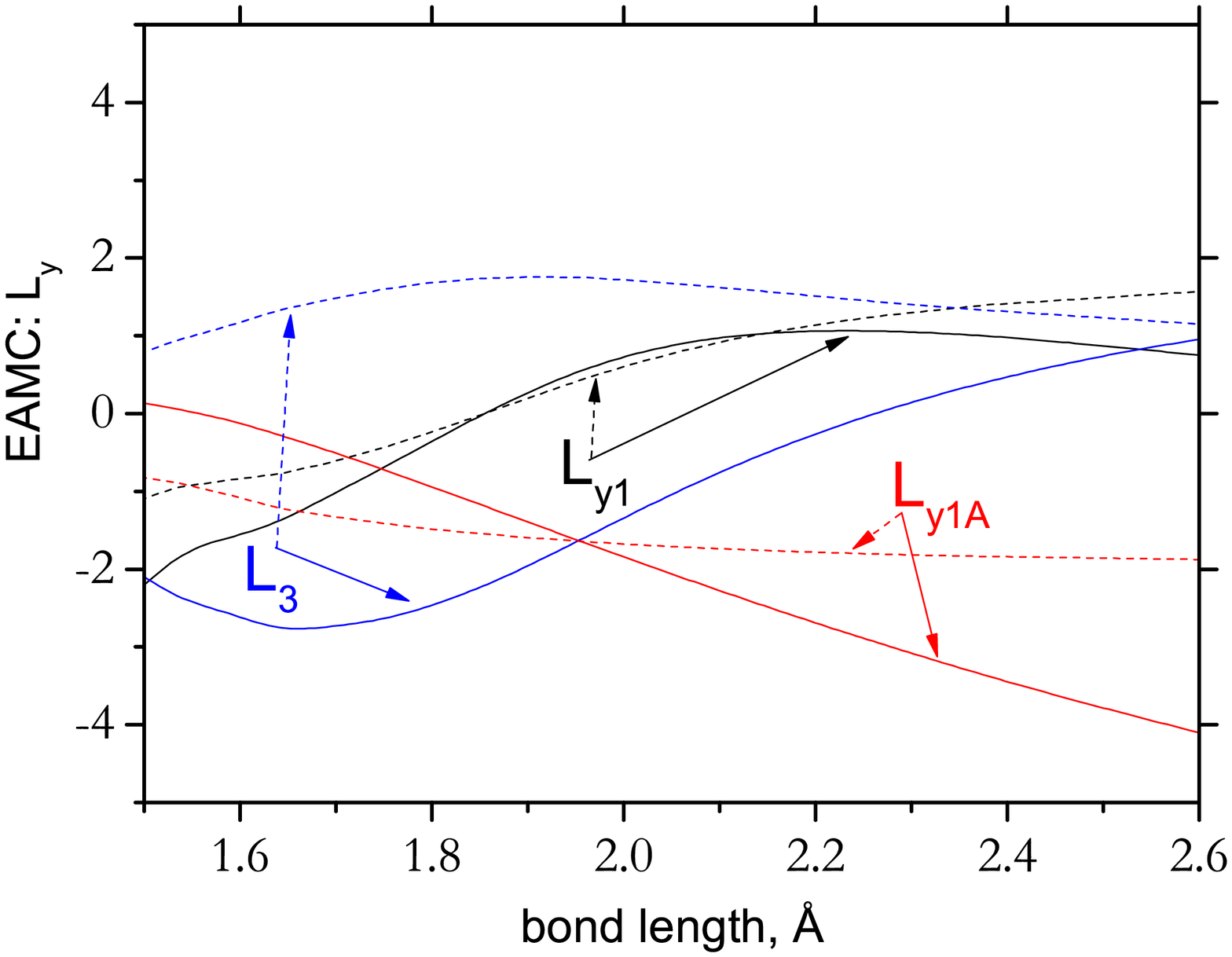}
\caption{
\label{f:SOCs}
{\it Ab initio}  (dotted curves) and fitted (full curves) spin-orbit (left) and
electronic angular momentum (right) coupling curves for the lowest five electronic
states of CaO, where SO$_1$($X$--$a$), SO$_2$($A'$--$b$), SO$_3$($a$--$A'$), SO$_4$($b$--$a$), SO$_{1A}$($A$--$a$),
 $L_{y1}$($X$--$A'$), $L_{y1A}$($A$--$A'$),  and $L_{y3}$($a$--$b$) are  defined in Eqs.~(10,24) of \citet{12KhLeBr.CaO}. }
\end{center}
\end{figure}

\begin{figure}
\begin{center}
\includegraphics[width=220pt]{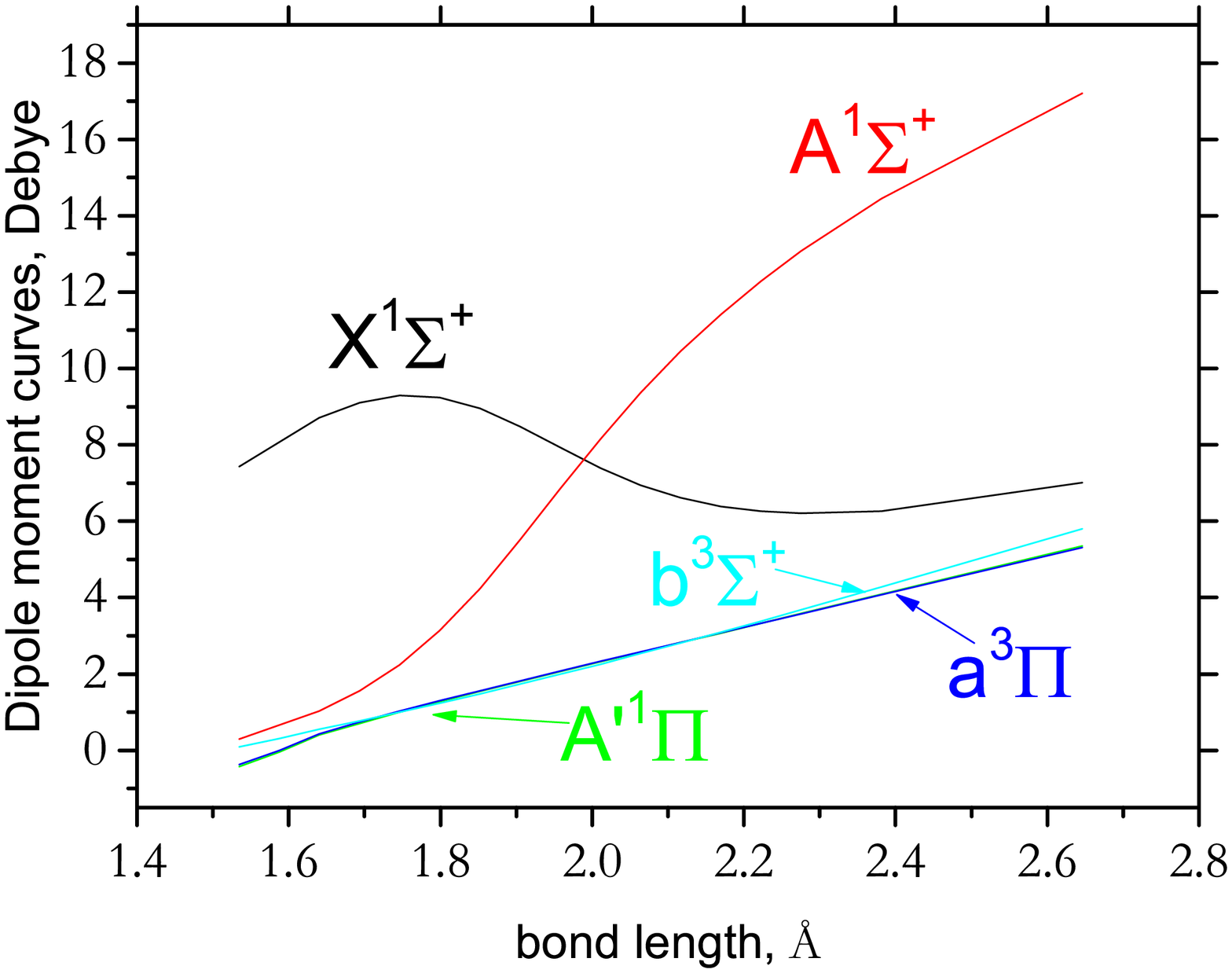}
\includegraphics[width=220pt]{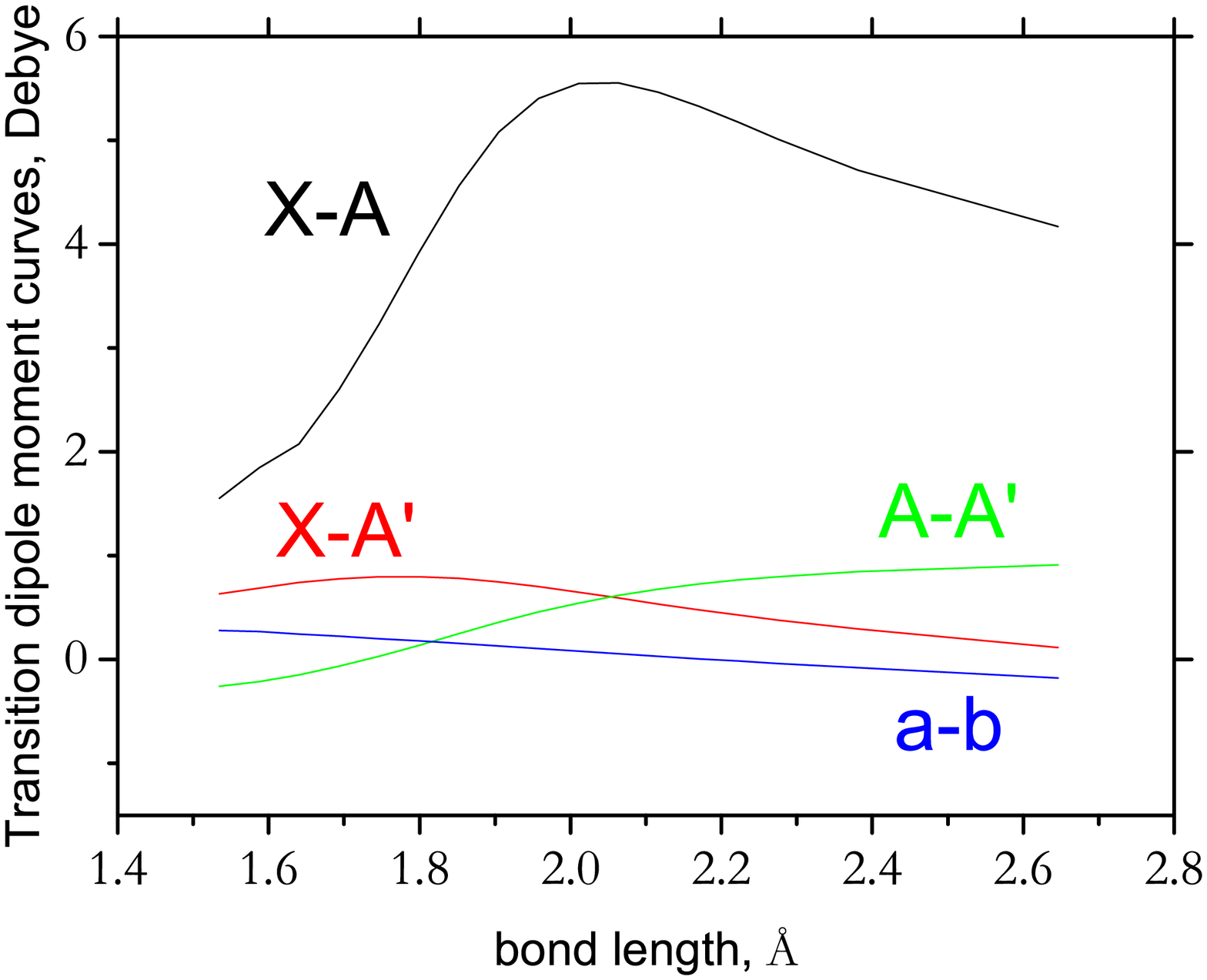}
\caption{
\label{f:DMC}
{\it Ab initio} dipole moment curves for the lowest five electronic
states of CaO \citep{11KhBrLe.CaO,12Lexxxx.CaO}: permanent dipoles (left); transition dipoles (right).}
\end{center}
\end{figure}

We use the program \duo\ \citep{jt609} to solve the fully coupled
Schr\"{o}dinger equation for the lowest five electronic states of CaO. The
details of the \duo\ methodology used for building accurate, empirical line
lists for diatomic molecules has been extensively discussed elsewhere
\citep{jt589,jt598,jt599,jt609,jtdiat}. The initial
PECs, SOCs, EAMCs used were taken from \citet{11KhBrLe.CaO,12KhLeBr.CaO,12Lexxxx.CaO} and are shown in Figs.~\ref{f:PECs}, \ref{f:SOCs}, and \ref{f:DMC}. These curves were obtained using the multi-reference configuration interaction (MRCI)
calculations in conjunction with the aug-cc-pV5Z and cc-pCV5Z basis sets for O and Ca,
respectively.

Our \duo\ calculations  used a grid-based Sinc method
with 501 points ranging from 1 to 4 \AA\ to solve the five un-coupled
vibrational eigen-problems for each state separately. The lowest 80 (\X) and
50 (\A, \Ap, \astate, and \bstate) vibrational eigenfunctions were then used as the vibrational
basis functions for the ro-vibronic, fully coupled problem in a Hund's case~a
representation. This problem is solved variational for each total angular momentum
quantum number, $J$, and parity by explicit diagonalisation of the coupled-states
Hamiltonian.

\section{Experimental data and refinement}

The \ai\ PECs, SOCs, and EAMs were refined by fitting to the experientially
derived energies and transition frequencies. This refinement used the same fitting approach as
\citet{jt598}.

Only the \astate\ state dissociates to lowest products of O($^3$P) and
Ca($^1$S); the other curves go asymptotically to the first excited
state of Ca, Ca($^3$P), although this also involves avoided crossings
in most cases. This asymptote lies 15200 \cm\ (1.88 eV) higher than
the lowest dissociation channel.  We used the experimental
dissociation value 4.11$\pm$ 0.07~eV of \cite{80IrDaxx.CaO} to set the
dissociation asymptote of the \astate\ relative to the minimum of the
ground electronic state, while the next dissociation channel O($^3$P)
+ Ca($^3$P) was fixed to the value $4.11+1.88 = 5.99$~eV. This value
can be compared to the $D_{\rm e}$ value of 6.13~eV (MRCI+Q) by
\citet{11KhBrLe.CaO}. The corresponding estimates of $D_{\rm
  e}$(O($^3$P) + Ca($^1$S)) from the literature are 4.049~eV
(CCSD(T)/cc-pCVQZ) by \citet{03IrOrMa.CaO}, 3.29~eV (MRCI +
Q)~\citep{11KhBrLe.CaO} 4.05~eV $\pm 0.07$ (empirical $D_0$) and
3.081~eV (deperturbative $D_{0}$) due to \citet{74Fixxxx.CaO}. Our
zero-point-energy is 364.675~\cm\ (=0.045~eV).

It is technically easier, at least initially, to do empirical
refinement by fitting to energies rather than transition frequencies.
To do this we built a set of experimental energies from different
sources as follows. The energy term values corresponding to the \Ap,
\astate, \bstate\ states were taken directly from
\citet{89NoCrSc.CaO,90BaFix1.CaO}.  Additional \X\ and \Ap\ energies
were derived from the transition frequencies reported by
\citet{00FoPoPi.CaO} (supplementary material) using combination
differences. We also initially used a large set of experimentally
derived term values of the \A\ ($v=0 - 5$), \astate\ ($v=6,9,12$) and
\Ap\ ($v=7, 13$) vibronic bands, all with $J\le 60$ by \citet{05VaTuFo.CaO}.  After a
preliminary refinement of the \ai\ curves to this set, our prediction
had significantly improved and we could confidently start working
directly with the experimental frequencies from the supplementary
material by \citet{05VaTuFo.CaO} (kindly provided by one of the
authors). This material is a more extensive set of actual
experimental emission frequencies
ranging from $J=0$ up to $J=118$, i.e. even beyond $J=60$ reported in their paper.
All transitions are assigned $J$ and parity
(rigorous quantum `numbers') as well as vibronic states labels (\A,
$v=0-5$). Some transitions from the resonance regions are represented
in this material by two or three entries with identical assignment.
These extra lines appear due to the intensity stealing caused by the
resonances of closely lying states (in this case highly vibrationally
excited \astate\ and \Ap). Despite the fact that resonances
make the modelling much more challenging, they provide a unique
opportunity of accessing energies from the dark electronic states,
which otherwise are unaccessible or difficult to observe.

Using these data and relying on van Groenendael et al's
rigourous quantum  numbers ($J$ and parity)
we have reconstructed the upper state energies associated with
these bands by combining their experimental \A\ --
\X\ transitions (including the resonance ones) with `our'
experimentally derived energies of the (lower) \X\ state.  Most of the
upper state energies were supported by 6 -- 10 transitions and all
energy levels characterised by only a single transitions were omitted
from our analysis, thus confirming the correct assignment of their $J$s and parities.
The \A\ state ro-vibronic energy levels were
obtained as an average of the corresponding upper states energy
levels, which resulted in $~630$ levels including $~$250 new ones in
addition to the set reported by \citet{05VaTuFo.CaO}. It should be
noted that we could not find any transitions associated with the
\astate\ ($v=6$) vibronic band as well with the $J=38,39$, $v=9$
(\astate) levels in these data.  Therefore they were not included into
our fitting set.

Last but not least, a set of experimentally derived energies
were extracted from the  experimental work by
\citet{50HuLaxx.CaO}, where the \A\ -- \X\ system was analysed. Most
of these transition wavenumbers, which are reported with an uncertainty
of $>$0.01~\cm, are outdated by the more accurate data of
\citet{05VaTuFo.CaO}, except those involving the $(6,3)$ and some of $(5,2)$
vibrational bands. Using these transitions together with `our'
experimentally derived \X-state energy levels we were able to extend
our fitting set by about 30 term values from the (\A, $v=6$) vibronic
band.

\begin{figure}
\begin{center}
\includegraphics[width=220pt]{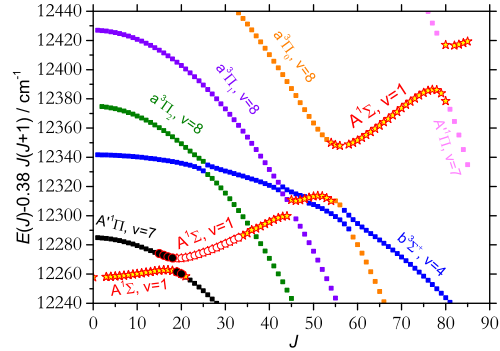}
\caption{
\label{f:A-X-1}
Reduced energy term values of CaO in the region of the \A\ ($v=1$) vibronic band: Filled and empty
circles represent the term values assigned in \protect\citet{05VaTuFo.CaO} to
\Ap\ ($v=7$) and \A\ ($v=1$), respectively.
Stars are the the new  term values assigned to \A\ ($v=1$) and
filled triangles are term values from \protect\citet{05VaTuFo.CaO} assigned here
to \Ap\ ($v=7$). Squares represent the calculated \duo\ levels. }
\end{center}
\end{figure}

The ro-vibronic assignment (both using rigourous and approximate
quantum labels) of the experimental energies/frequencies is crucial in
the fits. At the first step we had to use the assignment provided
by \citet{05VaTuFo.CaO}. However after initial fits we could already
rely on the assignment suggested by our \duo\ model, which is based on
the largest basis set contributions (i.e. largest expansion
coefficient). It turned out, however, that our assignments differ at high $J$
from those by \citet{05VaTuFo.CaO} in one important aspect
related to the behaviour of resonating vibronic states at and after
their crossing (in terms of $J$ progression). This is illustrated in
Fig.~\ref{f:A-X-1}, where a network of the rovibonic progression as
a function of $J$ from the region of the vibronic state \A\ ($v=1$) is
shown. The circles indicate the assignment by \citet{05VaTuFo.CaO}
(levels for $J=20-34$ assigned as (\Ap, $v=7$)) and squares represent
our assignment (all levels for $J=0-85$ assigned as (\A, $v=1$),
except five crossing states assigned as (\Ap, $v=7$)). In fact the latter
is more logical as the strongest transitions should belong to
the \A\ -- \X\ band across all $J$ except perhaps in the crossing
regions. This indicates the limitations of the effective rotational
Hamiltonian models, which most likely hindered van Groenendael et al's
analysis of this vibronic progression for $J>34$  and not reported in their paper.

%\red{I am not sure I can find l. 52: 5.      Pg. 4, l. 52. “and not reported in their paper.”}

\begin{figure}[htb!]
\begin{center}
\includegraphics[width=220pt]{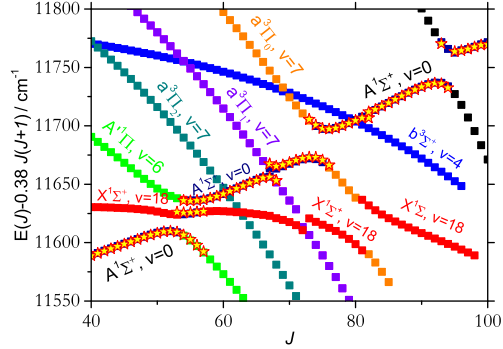}
\caption{
\label{f:A-X:v=0}
Reduced energy term values of CaO in the region of the \protect\A\ ($v=0$) vibronic band: Squares represent the calculated term values and stars show the experimental energies derived using the experimental frequency wavenumbers reported by \protect\citet{05VaTuFo.CaO} (new for $J>60$).}
\end{center}
\end{figure}

We also found similar cases where other vibronic
states cross the \A\ $v=0-5$ progressions: according to the effective
Hamiltonian results \citep{05VaTuFo.CaO} the \A\ progressions tend to
switch either to \astate\ or \Ap\ when they cross. This contrasts
with our picture, which places most of the transitions from this system
in the strong \A\ -- \X\ band, except for those resonance states that
appear to be due to the intensity stealing (\X, \astate, and \Ap). This also
applies to our newly derived energies above $J=60$, as can be seen in
Fig.~\ref{f:A-X:v=0}. Therefore in our fits are based on our
assignment which we believe to be reliable.

As mentioned above, the positive effect of the resonances is that they
provide access to other vibronic bands otherwise not visible to the
experiment. Especially valuable cases are when more than two states
cross which results in transitions from dark bands as well.  As an
illustration Fig.~\ref{f:A-X:v=0} shows the resonance region where
three vibronic states (\A, $v=1$), (\Ap, $v=6$) and \X\ $v=18$ meet at
$J=54-57$ and all three states are represented experimentally. The
theoretical data here represent our best model after the refinement.
As one can see the \duo\ calculations provide not just qualitatively
but also quantitatively correct description of the resonance. It is
reassuring that even the shapes of the $J$-progression of the both
dark vibronic states (\Ap, $v=6$, green squares) and (\X, $v=18$, red
squares) agree with the corresponding progressions formed by the
experimental levels (stars). Again, this is a unique situation when
due to intensity stealing we get access to a highly excited
(e.g. $v=18$)  region of the \X\ PEC; this is especially valuable
for high temperature applications.

Our complete energy set consists of 2204 term values
covering the rotational excitations up to $J=118$. At the final stage
of the refinement we could include the original experimental
transition wavenumbers from
\citet{50HuLaxx.CaO,05VaTuFo.CaO,00FoPoPi.CaO}. Such a combined
treatment was also used in our AlO work \citep{jt589}. Our final fit
optimised all \ai\ curves from \citet{11KhBrLe.CaO}, including five
PECs, six SOCs, and three EAMCs but excluding the electronic angular
momenta $L^2$ curves. The latter are effectively embedded into the
corresponding refined PECs. In the refinements of the \ai\ SOCs and EAMCs
we used the morphing procedure explained by \citet{jt589}. Most of
these curves remained very similar to the \ai\ ones (the morphing
factor is close to 1), except EAMCs $L_{y1A}$ and $L_{y3}$ (see
Fig.~\ref{f:SOCs}), which are scaled by the factors 0.13 and -0.36,
respectively.

The \bstate\ state has a very limited experimental description, only
the state $v=1$ with $J=14-25$ were characterised by
\citet{90BaFix1.CaO}, which is clearly not sufficient for an
unambiguous determination of the corresponding PEC. We have
represented the \ai\ PEC \citep{11KhBrLe.CaO} analytically using an
Extended Morse Oscillator (EMO) function \citep{EMO}, where the
dissociation value was changed to agree with the common value of
$D_{\rm e}$=5.99~eV (relative to \X) and the $T_{\rm e}$ and $r_{\rm e}$ values were
refined by keeping all other parameters from the Morse exponent
unchanged. Due to the limited data our results represent only a
possible solution for this state.

\begin{figure}
\begin{center}
\includegraphics[width=220pt]{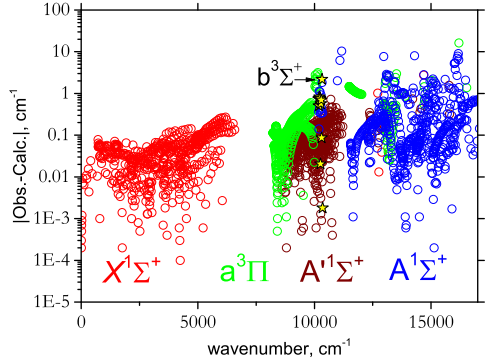}
\caption{
\label{f:rms}
Accuracy of the fit: The absolute values of the residuals between the experimentally derived (Obs.) and calculated (Calc.) energy term values for each electronic states are shown.  }
\end{center}
\end{figure}

Figure~\ref{f:rms} illustrates the accuracy of the fit.  The
root-mean-squares error for all experimental energies up to $J=118$ is
0.40~\cm, (0.07~\cm\ for \X, 0.15~\cm\ for \Ap, 0.43 for \astate,
0.8~for \bstate\ and 0.60~\cm\ for \A).

%\red{remove?}
%It should be noted that the perturbing \astate\ ($v=6$) energy levels (Table~1 by \citet{05VaTuFo.CaO}) were not used in the fit as not directly associated with any experimental transitions. Our predictions for these levels agree within x.x~\cm. \red{CHECK}

\begin{figure}
\begin{center}
\includegraphics[width=220pt]{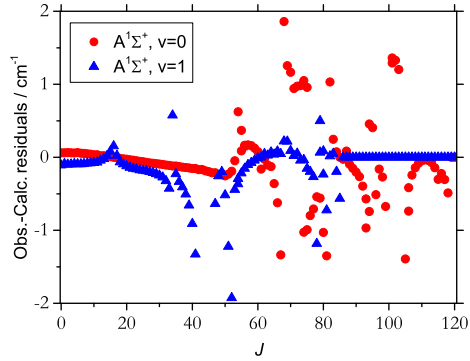}
 \caption{\label{f:A:v=0-1}
 Obs.-Calc. residuals for the  (\A, $v=0$) and (\A, $v=1$) ro-vibronic energy levels showing the influence of the resonances.}
\end{center}
\end{figure}

It should be noted that we could not resolve all resonances in the \A\
-- \X\ system, see Fig.~\ref{f:A:v=0-1}. In some cases the residuals in the
resonance region are up to 1--2 \cm.  We did not want to push the fit
too strongly by introducing more parameters because we were not completely
confident of the current representation of experimental data at the
high energy region. For example the \bstate\ state is one of the
important players (see Fig.~\ref{f:A-X:v=0}), but is clearly
under-sampled. Besides some of our objects already appear to be
over-fitted, which could affect the accuracy of prediction for high
$J$. Despite these relatively small numbers of outliers, we find that
the our model performs exceptionally well especially for \X, \Ap, and
\astate. Although our level of accuracy is not as high as that of
effective Hamiltonian models, our extrapolations for higher $v$ and
$J$ are likely to be much more reliable. However the high residuals in the
resonance regions indicate that more work is needed in the future.

We could also compare our theoretical energies to the `observed'
\astate\ ($v=6$) term values reported by \citet{05VaTuFo.CaO} (see
their Table~1) and omitted from our analysis due to absence of the
associated experimental evidence. We have found that our prediction is
systematically off the `observed' values of \citet{05VaTuFo.CaO} by
between 0.95 and 1.47~\cm, which could indicate artifacts in
our model due to resonances it does not fully account for.

Our assignments of the experimental states and experimental
transitions can be found in the supplementary material to this paper
as part of the complete list of experimental energies and frequency
wavenumbers used in the fit, together with the corresponding
obs.-calc. residuals. In some cases of strongly resonating states our
assignment (based on the largest contribution from the basis set
expansions) may become ambiguous due to equivalent contributions from
resonating basis set components and lead to, for example, duplicate
quantum labels. It can also fail for very high vibrational ($v\sim
~20$) and rotational ($J>100$) excitations when the basis set
functions are strongly mixed. Therefore in these and some cases the
assignment is not well defined and should be used with caution.

% figure
%

%For the final iteration we also complemented the energy set with
%3944 transition wavenumbers, mostly  $A$--$X$, by~\citet{05VaTuFo.CaO}
%covering $J\le 118$; some of these frequencies were excluded on the
%basis of ambiguous assignment.

%\red{COMPARE SPECTROSCOPIC CONSTANTS FROM 89NORMAN}.

%\red{TRY LAMBDA DOUBLING?}

\begin{table}
\caption{Sources and statistics of the experimental
data for CaO used in this work as well as the number of energy levels extracted and their ranges.
FTS=Fourier Transform Spectroscopy, LFS= Laser Fluorescence Spectroscopy, SDIF= Sub-Doppler intermodulated fluorescence, G= Grating}
\label{t:sources}
\begin{center}
\begin{tabular}{llrlclc}
%\hline\hline
%Transition moments                           &  This Work          &
%\citet{99ZeBlCh.AlO}$^{a}$     \\
\hline
\hline
Source                           &  Method          &    N of levels & State                       & $J$  &  $v$   &  Energy range, \cm         \\
\hline
%\citet{32Brxxxx.CaO}             &  G \red{???} &   --     &$ A{}^1\Sigma^{+}   $& 0--54    &  1 -- 4      &   --         \\
\citet{50HuLaxx.CaO}             &  G &   30  &$ A{}^1\Sigma^{+}   $& 2--119    &  0 -- 6      &   15995 -- 17264         \\
\citet{05VaTuFo.CaO}             &  FTS             &         210 &$ A{}^1\Sigma^{+}              $& 1--48    &  0 -- 5      &  11550--15824         \\
                                 &                  &         73 &$ a{}^3\Pi_{0}                 $& 2--60    &  6, 9,12          &  11488--15837        \\
                                 &                  &   32   &$ A\p{}^1\Pi                   $& 15--55   &  7,13             &  12365--16301        \\
\citet{00FoPoPi.CaO}$^{a}$             &  FTS             &         531 &$ A\p{}^1\Pi                   $& 1--78    &      0,1,2,3      &  8608--11219         \\
                                 &                  &         522 &$ X{}^1\Sigma^{+}              $& 0--77    &        0--7       &        0--6051          \\
\citet{89NoCrSc.CaO}             &  SDIF            &  414   &$ a{}^3\Pi                     $& 0--69    &        0--0       &  8349--9999        \\
\citet{90BaFix1.CaO}             &  LIF             &          12 &$ b{}^3\Sigma^{+}              $& 14--25   &                 1 &   10178--10354            \\
\citet{90BaFix1.CaO}             &  LIF             &          12 &$ a{}^3\Pi                $& 14--25   &                 3 &  10016--10164.1            \\
\hline
\hline
\end{tabular}
\mbox{}\\

$^{a}$ Derived from 3018 \A--\X\ transition frequencies using the combination
differences.

\end{center}

\end{table}

%\begin{equation}
% V(r)=V_{0}+B_{0}(\frac{r-r_{\rm e}}{r})^{2}\sum_{k=4}^{N}P_{k}(\frac{r-r_{\rm e}}{r})^{k-3}
%\label{eq:SPF}
%\end{equation}

% Ae=0.31957548085936E+05 a-state, Te=0.83866593563566E+04, De=

%For the spin-orbit and electronic angular momentum couplings from the
%corresponding  \ai\ curves  we employed the morphing approach
%described by  \citet{14PaHiTe.AlO}.

\section{Line list}

For the final line list production with \duo\ all potential energy and
coupling curves were mapped on a grid of 501  equidistant
points. This input is given in the supplementary material which allows
these curves to be extracted for other purposes than running \duo.

The line list was generated using the refined PECs, SOCa, EAMCs and \ai\
dipole and transition dipole moment curves. It contains  28~418~064   transitions
between states covering $J=0\ldots 221$ and the wavenumber range 0 -- 20000~\cm,
with the lower and upper state energies ranging up to 20000 and 35000~\cm,
respectively from the lowest five electronic states with the vibrational
excitations up to $v_{\rm} = 80$ for \X\ and 50 for the other states. For the sake of completeness,  the
partition function was evaluated using an energy set containing 130~660 levels covering
rotational excitations up to $J_{\rm max} = 400$.

The line list is divided into an energy file and a transitions file. This is
done using the standard ExoMol format \citep{jt548} based on the method
originally developed for the BT2 line list by \citet{jt378}. Extracts from
the CaO line list are given in Tables~\ref{tab:levels} and \ref{tab:trans}.
The full line list can be downloaded from the CDS, via
\url{ftp://cdsarc.u-strasbg.fr/pub/cats/J/MNRAS/xxx/yy}, or
\url{http://cdsarc.u-strasbg.fr/viz-bin/qcat?J/MNRAS//xxx/yy}.  The line
lists and partition function together with auxiliary data including the
potential parameters and dipole moment functions, as well as the absorption
spectrum given in cross section format \citep{jt542}, can all be obtained
also from \url{www.exomol.com}.

As a potentially more accurate alternative we provide an energy file where some
theoretical energies are replaced with their experimental counterparts, where available,
following a similar approach used by \citet{jt447,jt570} and \citet{jt615}.
By taking the advantage of the two-file structure of our line lists, this approach improves the frequencies without affecting the intensities. It guarantees the exact reproduction of the experimental frequencies (within the experimental error) when both upper and lower state energies are replaced. However a potential drawback is the loss of the consistency of the data due to the lack of the experimental information. We leave the choice between the pure theoretical and the hybrid energy files to the user.

\subsection{Partition function}

\begin{figure}
\begin{center}
\includegraphics[width=220pt]{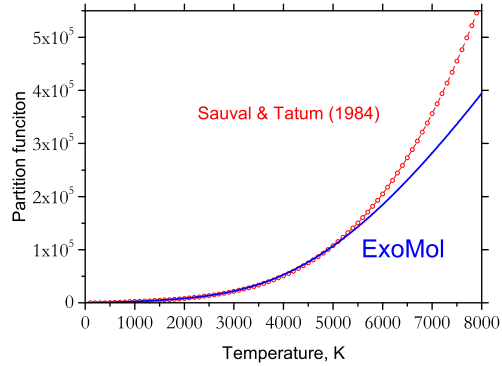}
\caption{Partition functions of CaO as a function of temperature. }
\label{f:pf}
\end{center}
\end{figure}

We used the computed energies to generate partition function values of CaO for
a large range of temperatures. Figure~\ref{f:pf} compares our partition function with that of
\citet{84SaTaxx.partfunc}. The agreement below 5000~K is very good. The differences above
this temperature may well be due to our neglect of higher-lying electronic states, although
assumptions in the methodology used by \citet{84SaTaxx.partfunc} may also contribute.
Similarly, we obtain very good agreement with the partition function given by CDMS (not shown)
\citep{cdms}, which only considers temperatures up to 500~K.

Following \citet{jt263} we represent our partition function using the
functional form
\begin{equation}
 \log_{10}{Q(T)}=\sum_{n=0}^{10} a_{n} [\log_{10}T]^{n},
\label{eq:part}
\end{equation}
where the fitting parameters $a_{n}$ are given in Table~\ref{t:pf}, which
reproduce the partition function in Fig.~\ref{f:pf} for entire region below
$8000$~K with the relative root-mean-square (rms) error of 3.6~\%.

\subsection{Examples of spectra}

\begin{figure}
\begin{center}
\includegraphics[width=220pt]{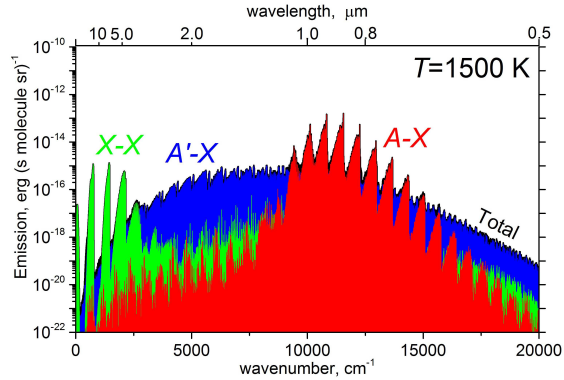}
\caption{Emission ($T=1500$~K) cross-sections (Gaussian profile, HWHM=1~\cm) of CaO showing the three strongest dipole allowed electronic bands of CaO. The total spectrum is in the background as indicated by the black out-line. }
\label{f:bands}
\end{center}
\end{figure}

\begin{figure}
\begin{center}
\includegraphics[width=220pt]{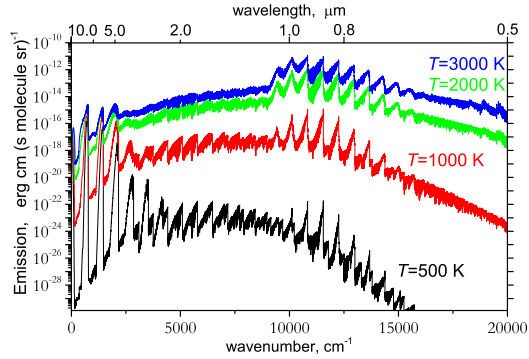} \\
\caption{Temperature dependence of the emission spectra of CaO (Gaussian profile, HWHM=1~\cm). }
\label{f:T}
\end{center}
\end{figure}

Figure~\ref{f:bands} gives an  overview of the CaO spectrum
and illustrates the contributions from each of the five
bands. A notable feature of this figure is strength of
all the bands in the system: several orders of magnitude stronger
than is typical of molecular transition intensities. This can
be attributed to the large charge separation in the \X\ state of CaO,
which is often represented by chemists as Ca$^{2+}$O$^{2-}$, the
variation of this separation with geometry and the different nature
of the excited states of CaO. All this means that the dipoles, which
are shown in Fig.~\ref{f:DMC}, are all large.
Figure~\ref{f:T} illustrates the strong temperature dependence of the spectra of
molecular CaO.

\begin{figure}
\begin{center}
\includegraphics[width=220pt]{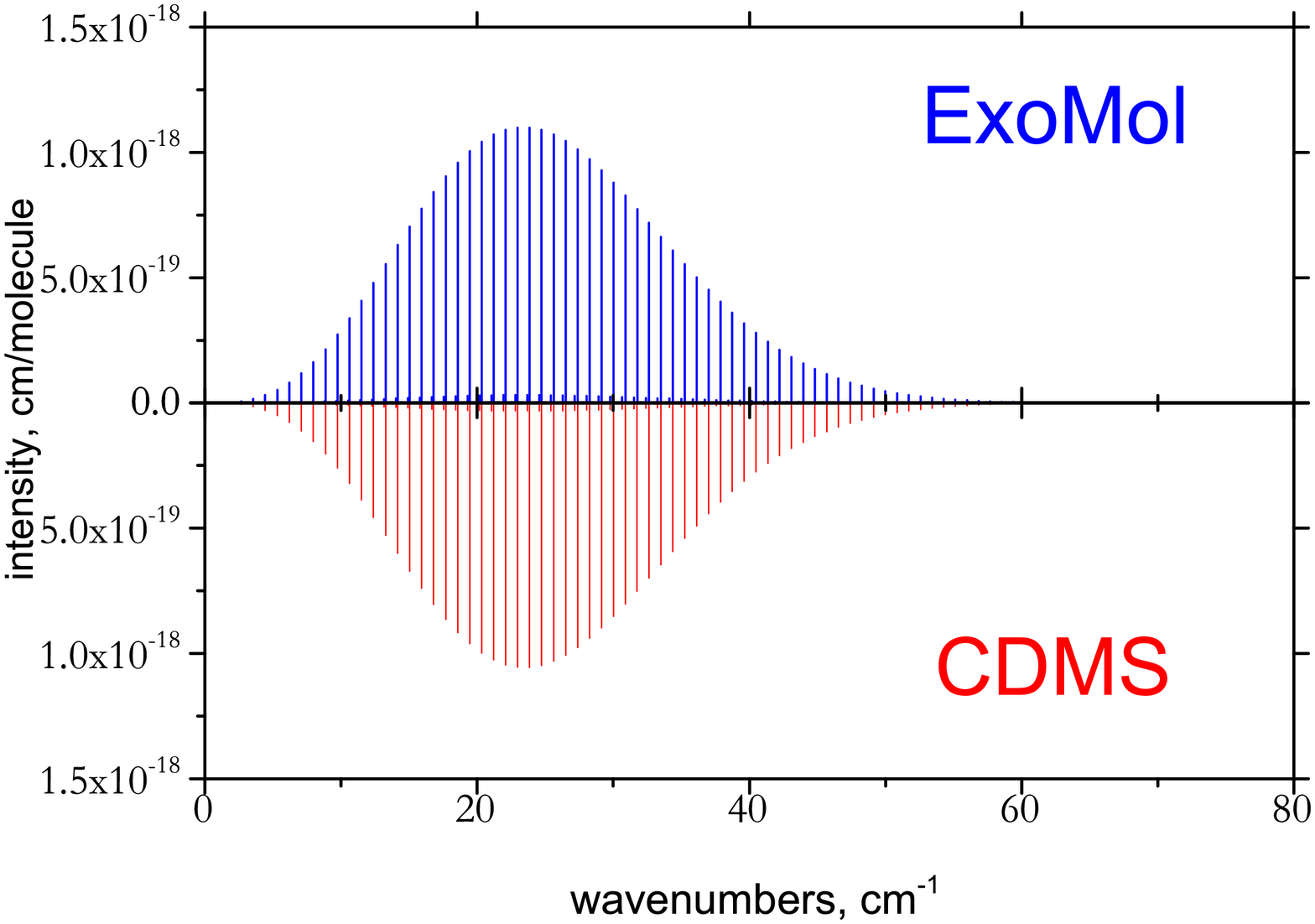}
\caption{Comparison with the CDMS rotational band at $T$=298~K. }
\label{f:CDMS}
\end{center}
\end{figure}

In order to test the quality of our theoretical line list, we
present a number of comparisons with previous works.  Figure~\ref{f:CDMS} compares a
rotational spectrum of CaO at $T=298$~K computed using our line list with the spectrum given in the CDMS
database \citep{cdms}. The agreement is very good: our spectrum is slightly stronger
which is a reflection of the slightly larger equilibrium \X\ state dipole
moment predicted by \citet{12KhLeBr.CaO} than the (unpublished) calculations
used by CDMS.

\begin{figure}
\begin{center}
\includegraphics[width=220pt]{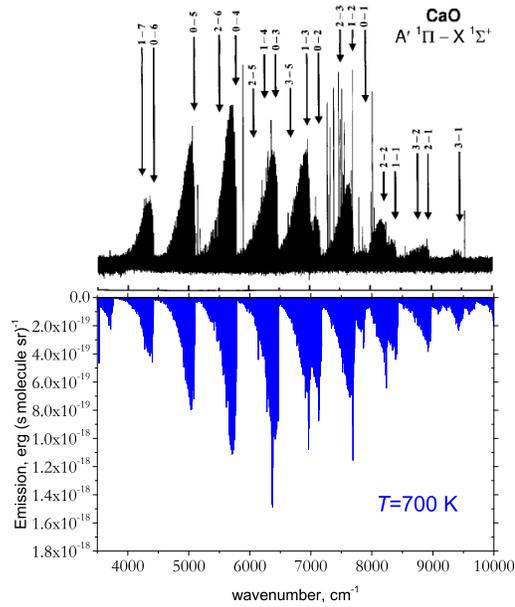}
\caption{Comparison with experiment \citep{00FoPoPi.CaO}: $T=700$~K emission with the Doppler profile. }
\label{f:exp1}
\end{center}
\end{figure}

\begin{figure}
\begin{center}
\includegraphics[width=220pt]{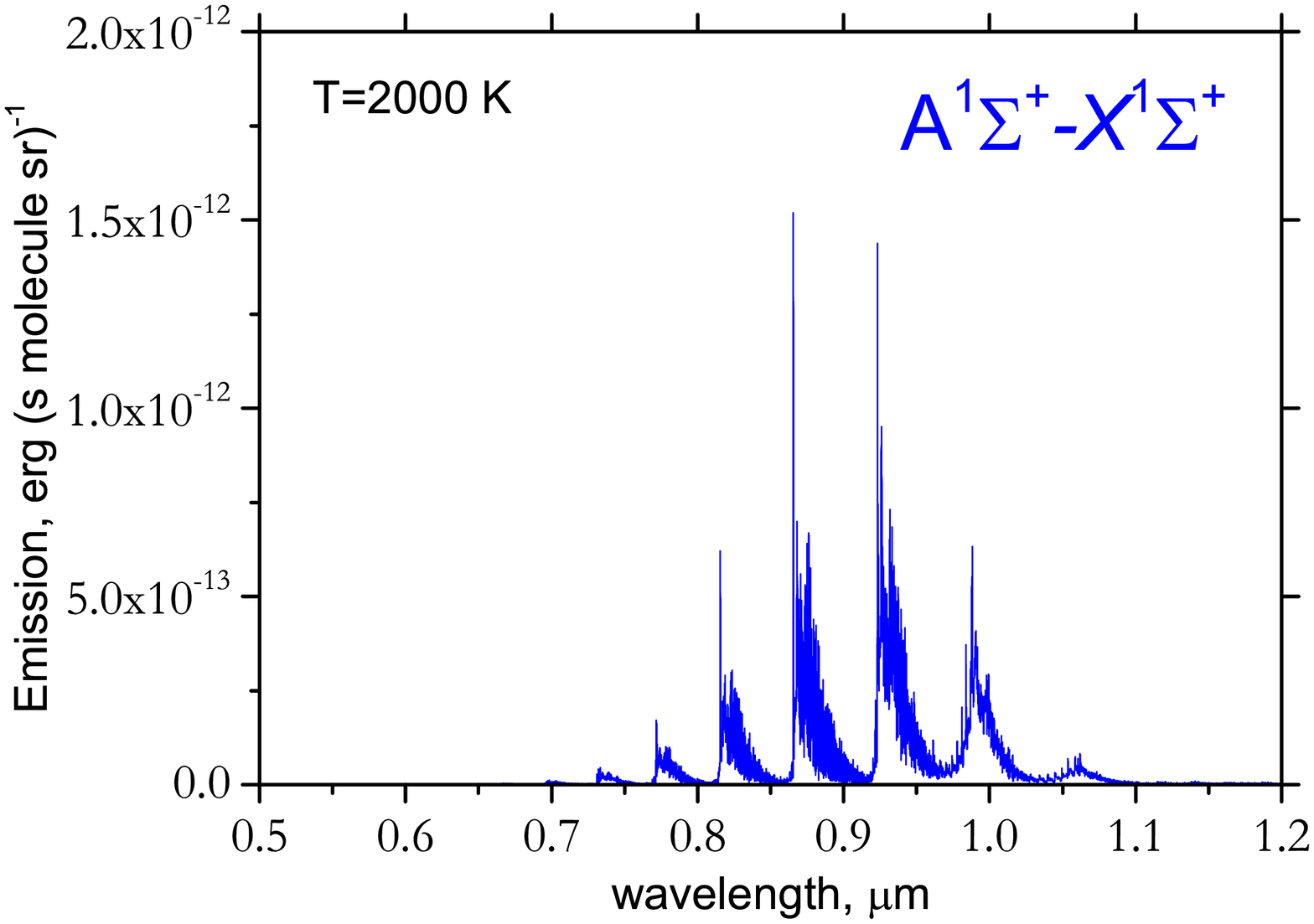}
\caption{Emission spectrum of CaO at $T=2000$~K, \A--\X\ band (Gaussian profile, HWHM=1~\cm). }
\label{f:2000K:A-X}
\end{center}
\end{figure}

Figure~\ref{f:exp1} compares an  \Ap\
-- \X\ band emission spectrum, simulated here and from the experiment by \citet{00FoPoPi.CaO}. Figure~\ref{f:2000K:A-X} illustrates the emission spectrum of \A--\X\ band of CaO simulated at $T=2000$~K using a Gaussian line profile with the half width at half maximum (HWHM) of 1~\cm.

\subsection{Lifetimes}

Excited-state lifetimes, where measured, provide an independent check on our Einstein  coefficients.
\citet{91PlNixx.CaO} measured an average emission lifetime for the  (\A,
$v=6$) vibronic state as $\tau=149 \pm 11$~ns for an unspecified distribution of $J$ levels.
We obtain $\tau=111$~ns  for the ($J=20$, \A, $v=6$) and $\tau=137$~ns for the ($J=40$, \A,
$v=6$) levels. Given the difficulty of making a precise comparison, this represents good agreement.

%\begin{figure}
%\begin{center}
%\includegraphics[width=220pt]{450-800nm.eps}
%\caption{Comparison with experiment \citep{75FiCaJo.CaO}: $T=3000$~K emission with the Gaussian (HWHM=1~\cm) profile. \red{SOMETHING IS NOT RIGHT} }
%\label{f:exp1}
%\end{center}
%\end{figure}

\begin{table*}
\caption{ Partition function parameters (with $T$ in K) for CaO, see Eq.~(\ref{eq:part}).}
\label{t:pf}
\begin{center}
\begin{tabular}{lr}
\hline\hline
Parameter	& Value  \\
\hline
$a_0$   & -3.7660464541     \\
$a_1$   & 35.3884754545     \\
$a_2$   & -117.5161774390   \\
$a_3$   & 215.9129214690    \\
$a_4$   & -241.0345706670   \\
$a_5$   & 173.0826753920    \\
$a_6$   & -81.7063206245    \\
$a_7$   & 25.2070167715     \\
$a_8$   & -4.8895914087    \\
$a_9$   & 0.5408689698      \\
$a_{10}$  -0.0260030674       \\
\hline
\end{tabular}
\end{center}
\end{table*}

\begin{table*}
\caption{   Extract from the state file for CaO. Full table is available from
http://cdsarc.u-strasbg.fr/cgi-bin/VizieR?-source=J/MNRAS/xxx/yy.}
\label{tab:levels}
\begin{center}
\footnotesize
\tabcolsep=5pt
\begin{tabular}{rrrrrrlrrrr c rrr}
\hline
     $i$ & \multicolumn{1}{c}{$\tilde{E}$} &  $g$    & $J$  &  \multicolumn{1}{c}{$+/-$} &  \multicolumn{1}{c}{$e/f$} & State & $v$    & $\Lambda$& $\Sigma$ & $\Omega$& \\
\hline
       1  &        0.000000  &      1  &      0  &     +    &     f   &\verb!  X1Sigma+     !&      0  &       0  &       0  &       0     \\
       2  &      722.452270  &      1  &      0  &     +    &     f   &\verb!  X1Sigma+     !&      1  &       0  &       0  &       0     \\
       3  &     1435.257802  &      1  &      0  &     +    &     f   &\verb!  X1Sigma+     !&      2  &       0  &       0  &       0     \\
       4  &     2138.509986  &      1  &      0  &     +    &     f   &\verb!  X1Sigma+     !&      3  &       0  &       0  &       0     \\
       5  &     2832.313497  &      1  &      0  &     +    &     f   &\verb!  X1Sigma+     !&      4  &       0  &       0  &       0     \\
       6  &     3516.781758  &      1  &      0  &     +    &     f   &\verb!  X1Sigma+     !&      5  &       0  &       0  &       0     \\
       7  &     4192.036586  &      1  &      0  &     +    &     f   &\verb!  X1Sigma+     !&      6  &       0  &       0  &       0     \\
       8  &     4858.206860  &      1  &      0  &     +    &     f   &\verb!  X1Sigma+     !&      7  &       0  &       0  &       0     \\
       9  &     5515.427402  &      1  &      0  &     +    &     f   &\verb!  X1Sigma+     !&      8  &       0  &       0  &       0     \\
      10  &     6163.838078  &      1  &      0  &     +    &     f   &\verb!  X1Sigma+     !&      9  &       0  &       0  &       0     \\
      11  &     6803.582280  &      1  &      0  &     +    &     f   &\verb!  X1Sigma+     !&     10  &       0  &       0  &       0     \\
      12  &     7434.805019  &      1  &      0  &     +    &     f   &\verb!  X1Sigma+     !&     11  &       0  &       0  &       0     \\
      13  &     8057.647658  &      1  &      0  &     +    &     f   &\verb!  X1Sigma+     !&     12  &       0  &       0  &       0     \\
      14  &     8352.058886  &      1  &      0  &     +    &     f   &\verb!  a3Pi         !&      0  &       1  &      -1  &       0     \\
      15  &     8672.234580  &      1  &      0  &     +    &     f   &\verb!  X1Sigma+     !&     13  &       0  &       0  &       0     \\
      16  &     8891.506388  &      1  &      0  &     +    &     f   &\verb!  a3Pi         !&      1  &       1  &      -1  &       0     \\
      17  &     9278.622853  &      1  &      0  &     +    &     f   &\verb!  X1Sigma+     !&     14  &       0  &       0  &       0     \\
      18  &     9424.671614  &      1  &      0  &     +    &     f   &\verb!  a3Pi         !&      2  &       1  &      -1  &       0     \\
      19  &     9876.485742  &      1  &      0  &     +    &     f   &\verb!  X1Sigma+     !&     15  &       0  &       0  &       0     \\
      20  &     9952.328458  &      1  &      0  &     +    &     f   &\verb!  a3Pi         !&      3  &       1  &      -1  &       0     \\
      21  &    10457.945342  &      1  &      0  &     +    &     f   &\verb!  X1Sigma+     !&     16  &       0  &       0  &       0     \\
      22  &    10482.722422  &      1  &      0  &     +    &     f   &\verb!  a3Pi         !&      4  &       1  &      -1  &       0     \\
      23  &    10983.490378  &      1  &      0  &     +    &     f   &\verb!  a3Pi         !&      5  &       1  &      -1  &       0     \\
      24  &    11055.566757  &      1  &      0  &     +    &     f   &\verb!  X1Sigma+     !&     17  &       0  &       0  &       0     \\
      25  &    11488.371123  &      1  &      0  &     +    &     f   &\verb!  a3Pi         !&      6  &       1  &      -1  &       0     \\
      26  &    11550.470240  &      1  &      0  &     +    &     f   &\verb!  A1Sigma+     !&      0  &       0  &       0  &       0     \\
      27  &    11630.899954  &      1  &      0  &     +    &     f   &\verb!  X1Sigma+     !&     18  &       0  &       0  &       0     \\
      28  &    11998.949002  &      1  &      0  &     +    &     f   &\verb!  a3Pi         !&      7  &       1  &      -1  &       0     \\
      29  &    12199.329740  &      1  &      0  &     +    &     f   &\verb!  X1Sigma+     !&     19  &       0  &       0  &       0     \\
      30  &    12257.881985  &      1  &      0  &     +    &     f   &\verb!  A1Sigma+     !&      1  &       0  &       0  &       0     \\
      31  &    12501.179666  &      1  &      0  &     +    &     f   &\verb!  a3Pi         !&      8  &       1  &      -1  &       0     \\
      32  &    12761.226866  &      1  &      0  &     +    &     f   &\verb!  X1Sigma+     !&     20  &       0  &       0  &       0     \\
      33  &    12960.097091  &      1  &      0  &     +    &     f   &\verb!  A1Sigma+     !&      2  &       0  &       0  &       0     \\
      34  &    13002.600726  &      1  &      0  &     +    &     f   &\verb!  a3Pi         !&      9  &       1  &      -1  &       0     \\
      35  &    13316.141667  &      1  &      0  &     +    &     f   &\verb!  X1Sigma+     !&     21  &       0  &       0  &       0     \\
      36  &    13484.091690  &      1  &      0  &     +    &     f   &\verb!  a3Pi         !&     10  &       1  &      -1  &       0     \\
      37  &    13675.102575  &      1  &      0  &     +    &     f   &\verb!  A1Sigma+     !&      3  &       0  &       0  &       0     \\
      38  &    13864.261534  &      1  &      0  &     +    &     f   &\verb!  X1Sigma+     !&     22  &       0  &       0  &       0     \\
\hline

\end{tabular}
\end{center}

\mbox{}\\
{\flushleft
$i$:   State counting number.     \\
$\tilde{E}$: State energy in \cm. \\
$g$: State degeneracy.            \\
$+/-$:   Total parity. \\
$e/f$:   rotationless-parity. \\
$v$:   State vibrational quantum number. \\
$\Lambda$:  Projection of the electronic angular momentum. \\
$\Sigma$:   Projection of the electronic spin. \\
$\Omega$:   $\Omega=\Lambda+\Sigma$, projection of the total angular
momentum.\\}

\end{table*}

\begin{table}
\caption{  Extract from the transitions file for CaO.
 Full table is available from
http://cdsarc.u-strasbg.fr/cgi-bin/VizieR?-source=J/MNRAS/xxx/yy. }
\label{tab:trans}
\begin{center}
\begin{tabular}{ccrr}
\hline\hline
$f$	&  $i$ 		& 		$A_{fi}$ &  $\tilde{\nu}_{fi}$\\
\hline
       10571    &        10884   &    9.5518E-06       &        120.241863      \\
       21053    &        21375   &    1.9515E-05       &        120.242886      \\
        8726    &         9672   &    1.8658E-04       &        120.243522      \\
       11655    &        11950   &    5.0065E-06       &        120.243733      \\
       93209    &        93967   &    5.7055E-03       &        120.244192      \\
        2228    &         3175   &    7.3226E-07       &        120.244564      \\
       46727    &        46432   &    1.0599E-04       &        120.244658      \\
       44436    &        44774   &    1.4626E-04       &        120.245583      \\
       29037    &        28723   &    1.8052E-04       &        120.245669      \\
        4458    &         4805   &    1.0431E-08       &        120.246396      \\
       69313    &        68434   &    5.0531E-06       &        120.248178      \\
       22640    &        22985   &    1.1281E-07       &        120.248891      \\
       57027    &        56721   &    7.1064E-06       &        120.250180      \\
       15224    &        15547   &    1.9505E-05       &        120.250711      \\
        6779    &         6477   &    5.0457E-04       &        120.251564      \\
       46942    &        47278   &    3.5870E-04       &        120.252807      \\
       18700    &        17749   &    1.8354E-03       &        120.255243      \\
       42535    &        42205   &    2.0975E-04       &        120.257381      \\
       86554    &        86842   &    6.9691E-05       &        120.259492      \\
       14843    &        15190   &    2.0376E-04       &        120.259971      \\
       49206    &        48911   &    8.7081E-05       &        120.260231      \\
       17728    &        18060   &    1.4899E-04       &        120.260340      \\
       31636    &        31977   &    8.7497E-04       &        120.260929      \\
  \hline
\end{tabular}

\noindent
 $f$: Upper  state counting number;
$i$:  Lower  state counting number; $A_{fi}$:  Einstein-A
coefficient in s$^{-1}$; $\tilde{\nu}_{fi}$: tranisiton wavenumber in \cm.

\end{center}
\end{table}

%some comparision with astronomy if available

\section{Conclusions}

We present a comprehensive line lists for calcium oxide, CaO.  This line list can be downloaded from the CDS, via
ftp://cdsarc.u-strasbg.fr/pub/cats/J/MNRAS/, or
http://cdsarc.u-strasbg.fr/viz-bin/qcat?J/MNRAS/, or from www.exomol.com.
CaO is one of a number of molecules which may be observable constituents of hot, rocky exoplanets.
The ExoMol project has already provided linelists for a number of other, similar molecules including
MgH and CaH \citep{jt529}, SiO \citep{jt563}, NaCl and KCl \citep{jt583}, AlO \citep{jt598}, ScH \citep{jt599}
 and NaH \citep{jt605}.
A number of other potentially important species,
including VO \citep{jt625}, TiO, TiH, NiH and CrH \citep{jtCrH}, are currently being studied. The ExoMol database is also currently being upgraded
to provide more extensive lists of molecular properties
including pressure broadening, lifetimes and Land\'e $g$-factors;
details of this will be given elsewhere \cite{jtexo}.

Using our refined model we could re-assign vibronic labels for a large
portion of the CaO transitions and levels given by \cite{05VaTuFo.CaO} for
$J\le 60$, and also assign their transitions with $J=60$ -- 118. This
is a reassuring illustration of the quality of our model and its ability to
correctly (at least in most cases) describe the perturbations of the
\A\ states due interactions with other vibronic bands.

\section{Acknowledgements}

We thank C. Leonard for the \ai\ curves and P. Bernath for the
experimental line list. This work was supported by the ERC under the
Advanced Investigator Project 267219 and made use of the DiRAC@Darwin,
DiRAC@COSMOS HPC cluster and Emerald CfI cluster. DiRAC is the UK HPC
facility for particle physics, astrophysics and cosmology and is
supported by STFC and BIS. The authors would like to acknowledge the
work presented here made use of the EMERALD High Performance Computing
facility provided via the Centre for Innovation (CfI). The CfI is
formed from the universities of Bristol, Oxford, Southampton and UCL
in partnership with STFC Rutherford Appleton Laboratory. We also
acknowledge the networking support by the COST Action CM1405 MOLIM.

\bibliographystyle{mn2e}
%\bibliography{journals_astro,jtj,AlO,methods,additional,linelists,CaO,ScH,diatomic,exoplanets}

\label{lastpage}

\end{document}